# Analysis of the most important variables which determine innovation among rural entrepreneurs

Author(s)*: Elena Lucia HARPA [1], Liviu MARIAN [2], Sorina MOICA [3], Iulia Elena APAVALOAIE [4]
Position: PhD Student[1], Prof., PhD[2], Lecturer, PhD[3], PhD Student [4]
University: Technical University of Cluj-Napoca[1,4]
"Petru Maior" University of Tirgu Mures[2,3]
Address: Cluj-Napoca, Memorandumului Str., No. 28, Romania
Email: elena_harpa@yahoo.com [1], liviu.marian@yahoo.com [2], sorina.moica@yahoo.com[3], iulia8@gmail.com [4]

## Abstract

***Purpose*** – The present research aims to highlight the main factors influencing the development of entrepreneurial innovation in a rural environment and to perform an empirical study with the purpose of assessing the main problems in rural development.

***Methodology/approach*** – The research performed is mostly of a quantitative nature, being based on the use of the questionnaire as research tool, although some of the questions were raised in order to collect respondents' impressions and opinions which would form the object fo qualitative research.

***Findings*** –The research outlines the fact that in the rural entrepreneurship innovation is performed with minimal investment in new technologies and depends on the entrepreneur's involvement in Innovation Systems Network. It was also noted that most of the entrepreneurs in rural environment are non-innovators.

***Research limitations/implications*** – The research question started to assess the key elements which identify the role of innovation among entrepreneurs in rural areas. Based on these facts, we determined the variables that make entrepreneurial innovation in rural areas, followed by the analysis of the most significant variable rural entrepreneurs in the Mures county.

***Practical implications*** – This result can support the creation of the future model of innovation in rural entrepreneurship.

***Originality/value*** – There are relatively few studies addressing the problem of research regarding innovation in rural areas. The emphasis is on national and regional studies, without differentiating between the rural and urban areas. Thus, the purpose of the analysis is to add a descriptive background related to the innovation at a micro-economic level in the rural areas.

## Introduction

The European Union is beginning to pay greater attention to the development of rural areas, providing more than simple agricultural support. With the growing popularity of the idea that starting and developing a business may constitute the best strategy for rural development, there are new reports from various international sources which are now questioning the advantages brought by innovation in entrepreneurship in order to support the economic development of rural areas. (OSLO Manual 2005).

## Innovation in rural entrepreneurship

This is how Wortman (1990) defined rural entrepreneurship: "rural entrepreneurship implies the creation of a new organization which introduces a new product, serves or creates a new market or uses a new technology in rural areas". This definition contains elements of innovation which can potentially affect a great part of the rural community in which the entrepreneurial activity will take place.

The concept of rural development constitutes a subject of continuous debate, especially with regard to the relative importance of its sectorial and territorial dimension. Rural development may be defined as "economic, social and environmental development as well as the development of administrations and territories characterized by low levels of population density, massive support as a form of subsistence agricultural land and non-urban settlements structure". (Fieldsend A., Kovács K., 2007).

The notion of rural development comprises all the actions directed towards improving the life quality of the populations in the rural areas, in order to preserve the natural and cultural landscape which is a guarantee of sustainable development of rural areas according to the conditions and specificity of the land. The rural development program may contain, according to conditions and necessities, the development of infrastructure, agriculture, tourism, small and medium-sized enterprises as well as the creation of jobs, but also ideas regarding the protection of the environment, education, community development (www.madr.ro).

At a certain extent, the economic objective of an entrepreneur and the social objective of rural development are much more interrelated in the rural environment than in the urban areas. For this reason, rural entrepreneurship is mainly community-based, having strong and extended family bonds and a relatively strong impact on the rural community (Hoy, Vaught ,1980).

The development of entrepreneurship means more than building a system of support for entrepreneurs. It is about creating entrepreneurial communities, changing the culture of rural areas and people so that they should adopt an entrepreneurial potential.

## Objectives, hypotheses, methodology

The main objective of this study is to analyze the level of innovation of rural entrepreneurs by analyzing the decision making strategies of small entrepreneurs. In order to fulfill this objective, we have set a series of specific research objectives:

O1. Identify the main elements regarding the role of innovation in rural entrepreneurship;
O2. Determine the variable elements composing entrepreneurial innovation in the rural environment;
O3. Analyze and interpret the most important variable elements identified in the Mures county.

In order to contribute to the fulfillment of the research objectives, we formulated the following research hypotheses whose test and ratification are detailed further on.

H1: The entrepreneur's level of innovation depends on his involvement in business networks.
H2: the level of innovation in rural environment is low because of the low level of EU grants and of the low investment capital for new technologies.

The quantitative analysis of the existing databases, both individually and in connection with the results of the survey, are intended to reveal interesting features regarding the factors determining the level of innovation in rural entrepreneurship. The survey contained information regarding the perception of the entrepreneurs on the various factors determining the innovation process, such as the development of the area, strategies of business development and the importance and relation with the business environment.

The questionnaire was distributed to a number of 500 entrepreneurs consisting of micro-enterprises, small and medium-sized enterprises as well as physical persons, individual businesses and family businesses from the Mures county. With a range of response of 47,4%, the final sample of the study contains 237 entrepreneurs.

If we compare the demographical date of the sampled businesses with the ones found in a statistical database, we find that they are comparable regarding the domain of activity, the level of employment, the years of functioning and juridical personality.

The questionnaire was grouped in 5 categories, according to the five essential elements which were identified: location, the entrepreneur's characteristics, the internal features of the business, the external features of the business and the characteristics of the innovation. Each of them also comprises variable elements which were included in the analysis.

Innovation, as the main element of the research, is described by four variable dependent components: types of innovation (process, product, marketing, organizational) brought to the development of the business, the degree of novelty implied by the innovation strategy, the entrepreneurship innovation level determined by the entrepreneur claims actions contained in the company's strategy and the level of excellence corresponding to the process of innovation generation and adoption by the entrepreneur.

**Determination of variables describing innovation in rural entrepreneurship**

Consequent to analyzing the main factors impacting the development of rural entrepreneurship in the Mures county, which was the object of previous personal research and after identifying the key-elements which would be analyzed (see figure 1), we designed a series of variables due to the multitude of the indicators to be measured and to their variety in the studied domains. These variables will be assessed by means of the questionnaire. The variables which were taken into account in this study were organized under the form of a conceptual model of analysis, as you can see in figure 2.

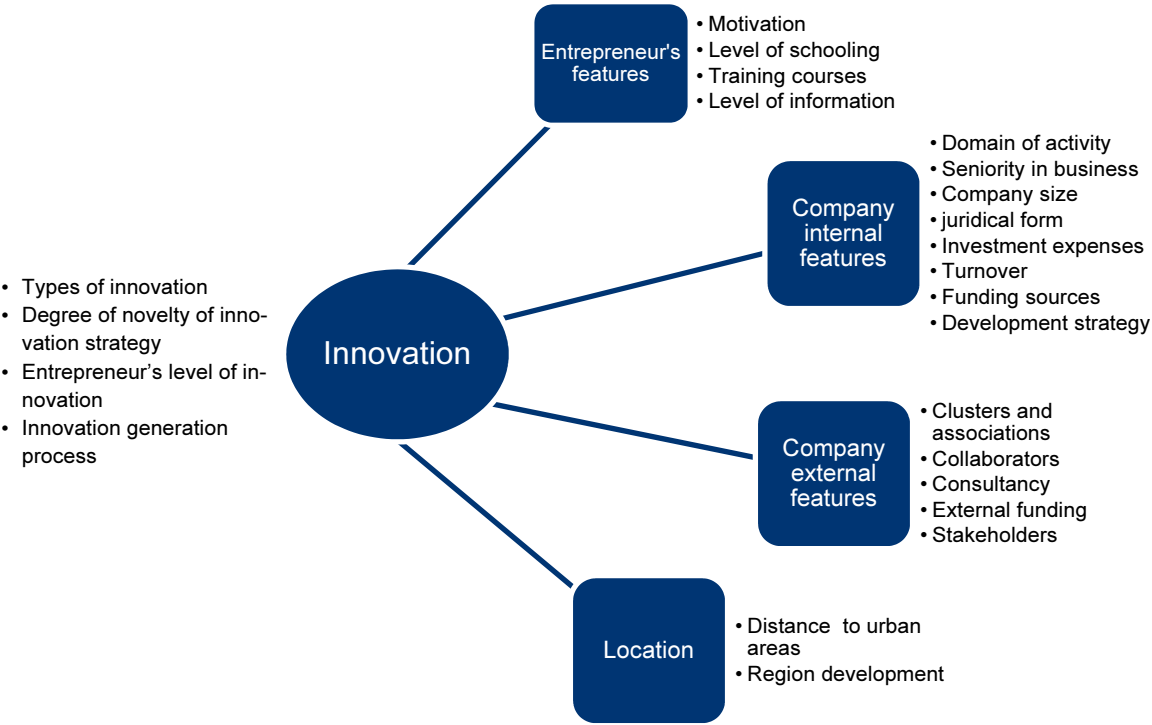

Figure 1: Conceptual analyze model

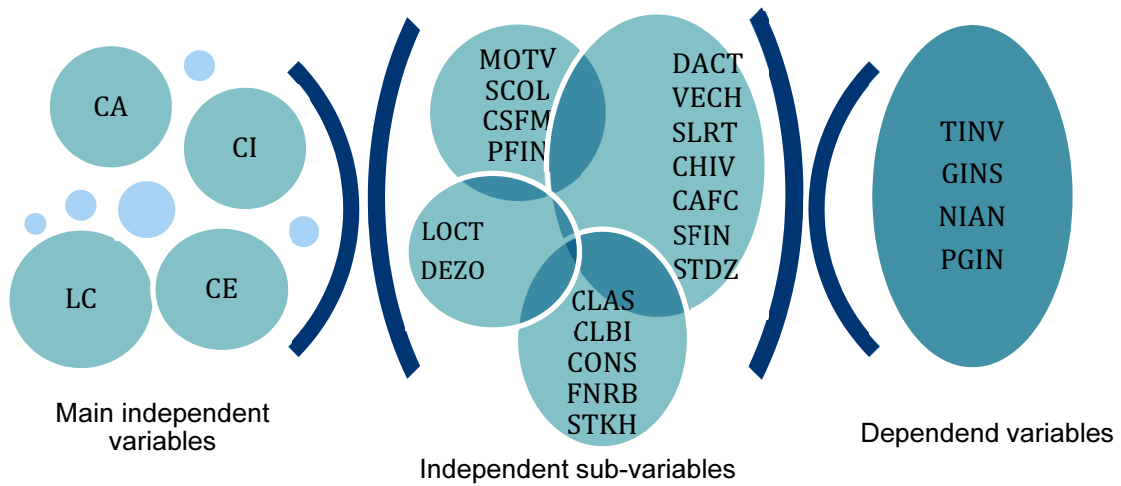

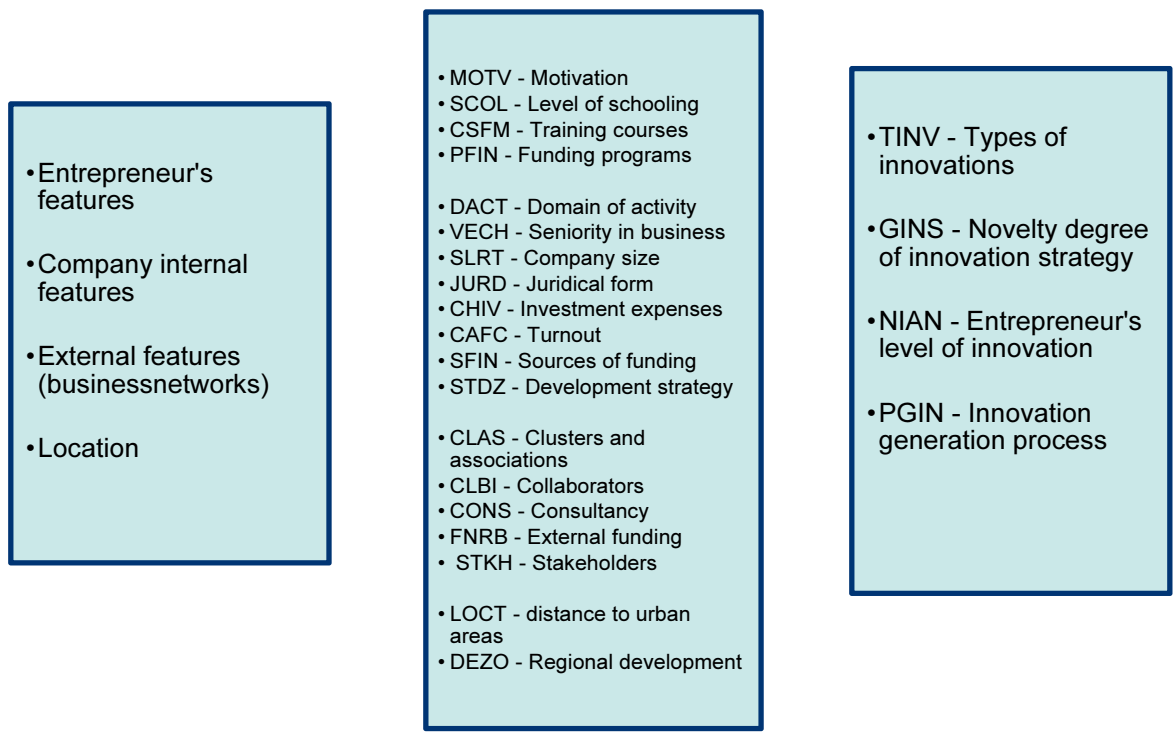

Figure 2 Variable analysis model

The conceptual model of analysis is composed of two types of variables. The independent variables are those with an independent behavior and which influence the dependent variables, who determining expected results in analysis. (Hernández, Fernández, Baptista, 2006).

The independent variables are grouped in four main categories of analysis: Location, Entrepreneur features, the internal features of the business and the external features of the business – the business networks. In their turn, these are formed of several variables which influence innovation, the main variable to be analyzed.

The dependent variables are the resulted variables. It is desired that their value is as high as possible. They are influenced at a greater or smaller extent by independent variables. In our case, we have four dependent variables, one of a quantitative type and three qualitative dependent variables by means of which we try to analyze the level of innovation in rural entrepreneurship.

## Process and interpretation of results

The first part of data analysis starts from descriptive statistics and contains a set of techniques / methods of organization and synthetic presentation of the collective data in order to shape an overall image regarding the main characteristics of the individual variables. One of the statistical methods which were used implies a descriptive analysis of the alternative variables in the study through the frequency analysis by using the mean score of influence (Moica 2013).

Tables 1 and 2 show the corresponding values of the dependent and independent variables analyzed in descriptive statistics and their review of maximal values which are far from the margins for the average confidence interval of 95%.

Table 1. Summary of descriptive statistics for independent variables

| Descriptive Statistics - independent variable | N Statistic | Minim Statistic | Maxim Statistic | Mean Statistic | Std. Error | Std. Deviation Statistic | Variance Statistic |
|---|---|---|---|---|---|---|---|
| Entrepreneur features | | | | | | | |
| Motivation | 237 | 11 | 19 | 15,02 | ,115 | 1,775 | 3,152 |
| Entrepreneur's schooling level | 237 | 2 | 7 | 4,57 | ,089 | 1,366 | 1,865 |
| Training courses | 237 | 0 | 1 | ,24 | ,028 | ,428 | ,183 |
| Funding programs | 237 | 0 | 1 | ,39 | ,032 | ,489 | ,239 |
| Company internal features | | | | | | | |
| Juridical form | 237 | 0 | 1 | ,49 | ,033 | ,501 | ,251 |
| Activity domain | 237 | 1 | 9 | 3,25 | ,147 | 2,271 | 5,156 |
| Company size | 237 | 0 | 243 | 8,58 | 1,460 | 22,484 | 505,533 |
| Financing sources for investment | 237 | 1 | 9 | 2,25 | ,083 | 1,284 | 1,648 |
| Development strategy | 237 | 3 | 9 | 5,62 | ,077 | 1,189 | 1,413 |
| Turnover | 237 | 2258 | 48011871 | 741500,41 | 233369,126 | 3592672,3 | 1290E+13 |
| Investment expenses | 237 | 0 | 3669252 | 74048,10 | 21630,24 | 332993,3 | 1108E+11 |
| Company external features | | | | | | | |
| Clusters and associations | 237 | 0 | 1 | ,35 | ,031 | ,477 | ,227 |
| Collaboration with institutions | 237 | 0 | 1 | ,09 | ,019 | ,291 | ,085 |
| Consultancy | 237 | 0 | 1 | ,20 | ,026 | ,403 | ,162 |
| External funding | 237 | 0 | 1 | ,21 | ,027 | ,409 | ,167 |
| Stakeholders | 237 | 14 | 25 | 19,62 | ,154 | 2,367 | 5,602 |
| Location | | | | | | | |
| Distance to urban areas | 237 | 1 | 4 | 1,93 | ,060 | ,923 | ,851 |
| Regional development | 237 | 14 | 23 | 18,88 | ,129 | 1,982 | 3,927 |

Table 2 - Summary of descriptive statistics for dependent variables

| Descriptive Statistics – Dependent variables | N Statistic | Minimum Statistic | Maximum Statistic | Mean Statistic | Std. Error | Std. Deviation Statistic | Variance Statistic |
|---|---|---|---|---|---|---|---|
| Innovation types | 237 | 0 | 5 | 1,91 | ,107 | 1,642 | 2,695 |
| Entrepreneur's level of innovation | 237 | 0,75 | 3,00 | 1,5591 | ,04390 | ,67585 | ,457 |
| Novelty degree of innovation strategy | 237 | 4 | 18 | 8,48 | ,249 | 3,840 | 14,743 |
| Innovation generation process | 237 | 1,8 | 7,4 | 4,513 | ,0865 | 1,3312 | 1,772 |

Regarding the profile of the entrepreneurs who are the object of the present sample, we analyzed a series of economic agents registered in the rural areas of the Mures county, from different sectors. 25,7% of the businesses are from agriculture, 23,6% from commerce and 21,2% from industry. According to the size of the business, in terms of the number of employees, almost 83% are micro-businesses, 33 are small businesses and 8 are medium businesses. From the point of view of the juridical status, 51% of the respondents have no juridical personality, being physical persons, individual businesses, family businesses, while a number of 116 respondents are trade companies. 21% of the entrepreneurs declared that they managed to attract EU funding and 90% of them resorted to a consultancy company.

Consequent to the application of the Parametric Correlation Analysis by means of the SPPS program, 7 new variables (main components) resulted. The number is too big for a rigorous interpretation of the information in the study. Analyzing the values of the correlation coefficients, it can be noted that there are both positive and negative correlations. Thus, it may be observed that:
- The entrepreneurs' level of innovation is positively correlated at an average intensity with the juridical status and the company seniority, but also with the degree of novelty implied by the innovation strategy.
- The consultancy regarding funding programs is averagely and directly correlated (0,455)
- The consultancy regarding external funding is directly and strongly correlated (0,717)
- External funding is averagely and directly correlated with the process of innovation generation (0,512)
- External funding is averagely and directly correlated with clusters and associations (0,407)

Following the second processing of the Parametric Correlation Analysis, by eliminating the seven variables mentioned above, resulted a number of 4 main components which explains 61,58% of the total variance of the 13 variables which remained in the study. The structure of each of the four components is rendered in table 3. From analyzing the structure of each component, it follows that:
- The first main component CP1 is formed of the following initial variables: consultancy, external funding, funding programs, innovation generation process and location;
- The second main component CP2 is formed of the following initial variables: turnover, company size and investment expenses;
- The third main component CP3 is formed of the following initial variables: company seniority, degree of innovation and the degree of novelty of the innovation strategy;
- The fourth main component CP4 is formed of the following initial variables: collaboration with institutions and clusters and associations.

Table 3. Total variance explained by the initial 13 variables remained in the study

| Variables | Component | | | |
|---|---|---|---|---|
| | CP1 | CP2 | CP3 | CP4 |
| Consultancy | ,818 | | | |
| External funding | ,789 | | | |
| Funding programs | ,725 | | | |
| Innovation generation process | ,691 | | | |
| Location | ,532 | | | |
| Turnover | | ,912 | | |
| Company size | | ,858 | | |
| Investment expenses | | ,563 | | |
| Company seniority | | | ,789 | |
| Entrepreneur's level of innovation | | | ,753 | |
| Novelty degree of innovation strategy | | | ,548 | |
| Collaboration with institutions | | | | ,863 |
| Clusters and associations | | | | ,739 |

*Extraction Method: Principal Component Analysis. Rotation Method: Varimax with Kaiser Normalization.a Rotation converged in 6 iterations.*

Following the non-hierarchical cluster analysis, one may note how the new main components are characterized by means of the 13 variables. In order to find the number of clusters, we first applied the non-hierarchical cluster analysis ith the Ward distance. Based on these data, 3 clusters were established. In order to see how the three clusters were formed on the 4 main components, we used the test k-means –cluster (Dixon 1992, Hitt, Hoskisson si Kim 1997). Thus:
- cluster C1 is defined by variables which form the main component C1, but it is not defined by the variables forming CP3.
- cluster C2 is defined by variables which form the main component CP3 but it is not defined by the variables forming CP1, therefore, clusters C1 and C2 are totally opposed.
- cluster C3 is defined by variables which form the main component CP2 but it is not defined by the variables forming CP3.

By analyzing the information in table 4, it may be noted that all the three clusters are statistically significant, according to the result of the ANOVA test. In order to obtain further details regarding the content of the three clusters according to the innovation indicators we applied the "crosstab" analysis between these variables (Sorensen & Stuart 2000). The results are presented in tables 5, 6 and 7.

Table 4. Results test ANOVA

| | Cluster | | Error | | F | Sig. |
|---|---|---|---|---|---|---|
| | Mean Square | df | Mean Square | df | | |
| Main component **CP1** | 24,192 | 2 | ,802 | 234 | 30,173 | **,000** |
| Main component **CP2** | 74,540 | 2 | ,371 | 234 | 200,673 | **,000** |
| Main component **CP3** | 51,229 | 2 | ,571 | 234 | 89,767 | **,000** |
| Main component **CP4** | 4,400 | 2 | ,971 | 234 | 4,531 | **,012** |

Table 5. Crosstab between the entrepreneur's level of innovation and the three clusters

| | **Cluster Number of Case** | | | Total |
|---|---|---|---|---|
| | 1 | 2 | 3 | |
| Non Innovators | 84 | 45 | | 129 |
| Innovators | 26 | 29 | 5 | 60 |
| High degree of Innovation | 8 | 39 | 1 | 48 |
| Total | 118 | 113 | 6 | 237 |

Table 6. Crosstab between the novelty degree of the innovation strategy and the three clusters

|  | Cluster Number of Case | | | Total |
|---|---|---|---|---|
|  | 1 | 2 | 3 |  |
| Non Innovators | 82 | 48 | 0 | 130 |
| Innovators | 27 | 35 | 4 | 66 |
| High degree of Innovation | 9 | 30 | 2 | 41 |
| Total | 118 | 113 | 6 | 237 |

Table 7 - Crosstab between the innovation generation process and the three clusters

|  | Cluster Number of Case | | | Total |
|---|---|---|---|---|
|  | 1 | 2 | 3 |  |
| Non Innovators | 21 | 55 | 0 | 76 |
| Innovators | 54 | 42 | 6 | 102 |
| High degree of Innovation | 43 | 16 | 0 | 59 |
| Total | 118 | 113 | 6 | 237 |

**Discussion and conclusions**

After processing the statistical data, it can be observed that the level of innovation of rural entrepreneurs is low. 54% of them fall into the category of Non-innovators and only 20% are characterized by high degrees of innovation. This result is mainly influenced by variables: consultancy, funding programs, external funding and investment expenses, which confirms our study hypothesis H2: the level of innovation in rural entrepreneurship is low due to the low level of EU fund attraction and of the low level of capital invested in new technologies.

The category of Non-innovators prevails in almost all the variables on the analyzed innovation. A higher presence of the second category of innovators is only noted through the analysis of the innovation generation process. This leads us to believe that rural entrepreneurs started actions of innovation generation, but, for various reasons, these entrepreneurs did not manage to complete the process. These reasons are explained by the entrepreneurs' lack of appurtenance to different clusters and associations and of collaboration with various institutions, which confirms hypothesis H1 of the study.

The research data confirm the low level of innovation in rural entrepreneurship as well as the lack of entrepreneurial culture and recommend an intensification of the efforts directed towards entrepreneurial education and stimulation of creativity.

**Acknowledgement**
This paper is supported by the Sectorial Operational Programme Human Resources Development POSDRU/159/1.5/S/137516 financed from the European Social Fund and by the Romanian Government